\documentclass{PoS}

\title{LHC signatures and cosmological implications of the E$_6$ inspired SUSY models}

\ShortTitle{LHC signatures and cosmological implications of the E$_6$ inspired SUSY models}

\author{\speaker{Roman Nevzorov}%
         \thanks{On leave of absence from the Theory Department, ITEP, Moscow, Russia.}\\
        University of Adelaide \\
        E-mail: \email{roman.nevzorov@adelaide.edu.au}}

\abstract{
The phenomenological implications of the $E_6$ inspired supersymmetric models
based on the Standard Model gauge group together with extra $U(1)_N$ gauge symmetry
under which right--handed neutrinos have zero charge are examined. In these models single
discrete $\tilde{Z}^{H}_2$ symmetry forbids the tree-level flavour changing processes
and the most dangerous operators that violate baryon and lepton numbers.
The two--loop renormalisation group flow of the gauge and Yukawa couplings is explored
and the qualitative pattern of the Higgs spectrum in the case of the quasi-fixed point scenario
is discussed. These $E_6$ inspired models contain two dark-matter candidates.
The presence of exotic states in these models gives rise to the nonstandard decays of the
lightest Higgs boson which are also considered.}

\FullConference{The European Physical Society Conference on High Energy Physics\\
                 22-29 July 2015\\
                 Vienna, Austria}

\begin{document}

\section{E$_6$ inspired SUSY models with extra $U(1)_{N}$ gauge symmetry}

It is well known that Supersymmetric (SUSY) Grand Unified Theories (GUTs) can lead to the
$U(1)$ extensions of the Minimal Supersymmetric Standard Model (MSSM), i.e. SUSY models
based on the Standard Model (SM) gauge group together with extra $U(1)$ gauge
symmetry. In particular, near the GUT scale $E_6$ can be broken to
$SU(3)_C\times SU(2)_W\times U(1)_Y\times U(1)_{N}\times Z_{2}^{M}$ group
\cite{Nevzorov:2012hs} where $Z_{2}^{M}=(-1)^{3(B-L)}$ is a matter parity and
\begin{equation}
U(1)_N=\frac{1}{4} U(1)_{\chi}+\frac{\sqrt{15}}{4} U(1)_{\psi}\,.
\label{1}
\end{equation}
Two $U(1)_{\psi}$ and $U(1)_{\chi}$ symmetries can originate from
breakings $E_6\to SO(10)\times U(1)_{\psi}$,
$SO(10)\to SU(5)\times U(1)_{\chi}$ \cite{King:2005jy}. To ensure anomaly cancellation the
low energy matter content of the $E_6$ inspired SUSY models with extra $U(1)_{N}$ gauge
symmetry is extended to fill out three complete $27$ representations of $E_6$
\cite{King:2005jy}--\cite{King:2005my}. Each $27_i$ multiplet contains SM family of quarks and leptons,
right--handed neutrino $N^c_i$, SM singlet field $S_i$ which carry non--zero $U(1)_{N}$ charge,
a pair of $SU(2)_W$--doublets $H^d_{i}$ and $H^u_{i}$, which have the quantum numbers of
Higgs doublets, and charged $\pm 1/3$ coloured triplets of exotic quarks $D_i$ and $\bar{D}_i$.
In addition to the complete $27_i$ multiplets the low energy particle spectrum is supplemented
by $SU(2)_W$ doublets $L_4$ and $\overline{L}_4$ from extra $27'$ and $\overline{27'}$ to
preserve gauge coupling unification \cite{King:2007uj}. Since in these models $N^c_i$ do not
participate in the gauge interactions these states are expected to gain masses at some intermediate scale,
shedding light on the origin of the mass hierarchy in the lepton sector and providing a mechanism
for the generation of the baryon asymmetry in the Universe via leptogenesis \cite{King:2008qb}.
The remaining matter supermultiplets survive down to the TeV scale. Different phenomenological
implications of the $E_6$ inspired SUSY models with extra $U(1)_{N}$ gauge symmetry were
considered in \cite{Nevzorov:2012hs}--\cite{Athron:2014pua}. Recently the particle spectrum
and the corresponding collider signatures were analysed within the constrained version of
this $U(1)_{N}$ SUSY extension of the SM \cite{Athron:2009bs}--\cite{Athron:2012sq}.

The presence of exotic matter in the $E_6$ inspired SUSY models lead to non--diagonal flavour
transitions and rapid proton decay. In order to suppress flavour changing processes
as well as baryon and lepton number violating operators one can impose a $\tilde{Z}^{H}_2$ symmetry.
Under this symmetry all superfields except $L_4$, $\overline{L}_4$, one pair of $H^{u}_{i}$ and
$H^{d}_{i}$ (i.e. $H_{u}$ and $H_{d}$) and one of the SM-type singlet superfields $S_i$ (i.e. $S$)
are odd. The $\tilde{Z}^{H}_2$ symmetry reduces the structure of the Yukawa interactions to
\begin{equation}
\begin{array}{c}
W = \lambda S (H_u H_d) + \lambda_{\alpha\beta} S (H^d_{\alpha} H^u_{\beta})+\kappa_{ij} S (D_{i} \overline{D}_{j})
+ \tilde{f}_{\alpha\beta} S_{\alpha} (H^d_{\beta} H_u) +f_{\alpha\beta} S_{\alpha} (H_d H^u_{\beta}) \\
+ g_{ij} (Q_i L_4) \overline{D}_j + h_{i\alpha} e^c_{i} (H^d_{\alpha} L_4) + \mu_L L_4\overline{L}_4 + W_{MSSM}(\mu=0)\,,
\end{array}
\label{2}
\end{equation}
where $\alpha,\,\beta=1,2$ and $i,\,j=1,2,3$.
At low energies the superfields $H_u$, $H_d$ and $S$ play
the role of Higgs fields. The vacuum expectation value (VEV) of $\langle S \rangle = s/\sqrt{2}$
breaks the extra $U(1)_N$ symmetry providing an effective $\mu$ term as well as the masses
of the exotic fermions and the $Z'$ boson ($M_{Z'}$). The VEVs of the $SU(2)_W$ doublets
$\langle H_d \rangle = v_1/\sqrt{2}$ and $\langle H_u \rangle = v_2/\sqrt{2}$ result in the
electroweak (EW) symmetry breaking (EWSB), inducing the masses of quarks and leptons.

\section{Quasi-fixed points and spectrum of Higgs bosons}

The superpotential (\ref{2}) involves a lot of new Yukawa couplings. To simplify our analysis of the
renormalisation group (RG) flow of the gauge and Yukawa couplings we assume that all Yukawa
couplings except $\lambda$ and the top--quark Yukawa coupling $h_t$ are sufficiently small and
can be neglected in the leading approximation. Then the superpotential (\ref{2}) reduces to
\begin{equation}
W\approx \lambda S(H_{d} H_{u})+h_t (H_{u} Q_3) u^c_3\,.
\label{3}
\end{equation}
For the purposes of RG analysis, it is convenient to introduce $\rho_t=h_t^2/g_3^2$ and
$\rho_{\lambda}=\lambda^2/g_3^2$, where $g_3$ is a strong gauge coupling.
\begin{figure}
\includegraphics[width=0.5\textwidth]{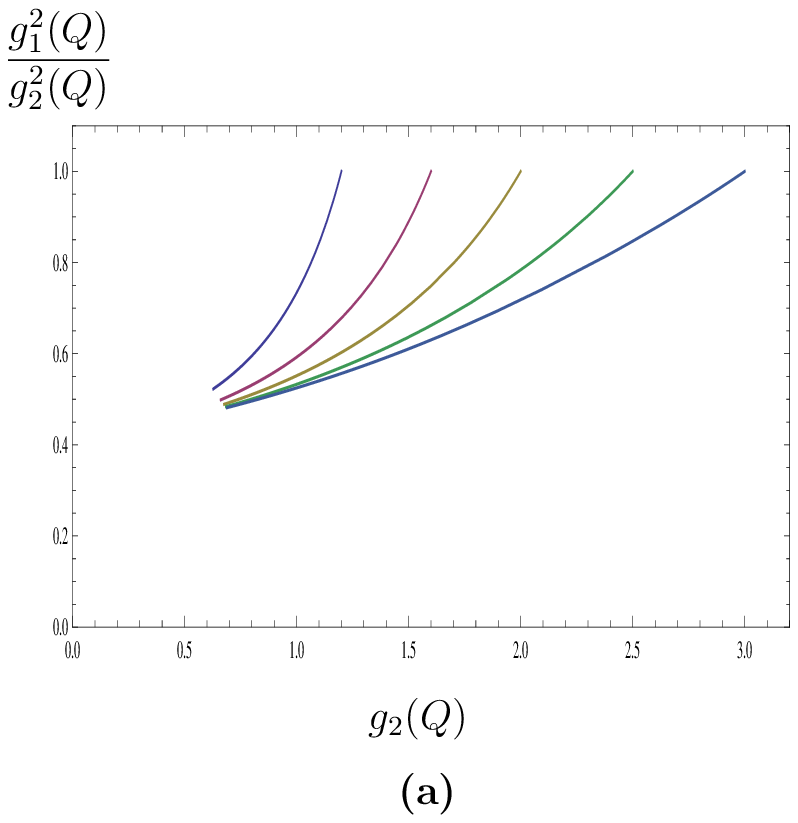}
\includegraphics[width=0.5\textwidth]{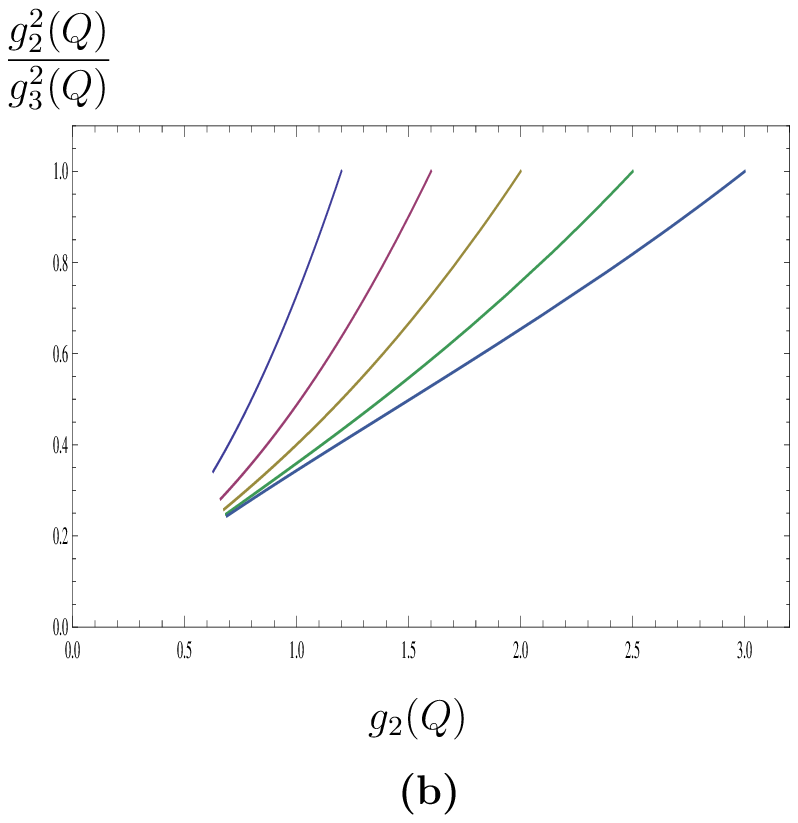}\\
\includegraphics[width=0.5\textwidth]{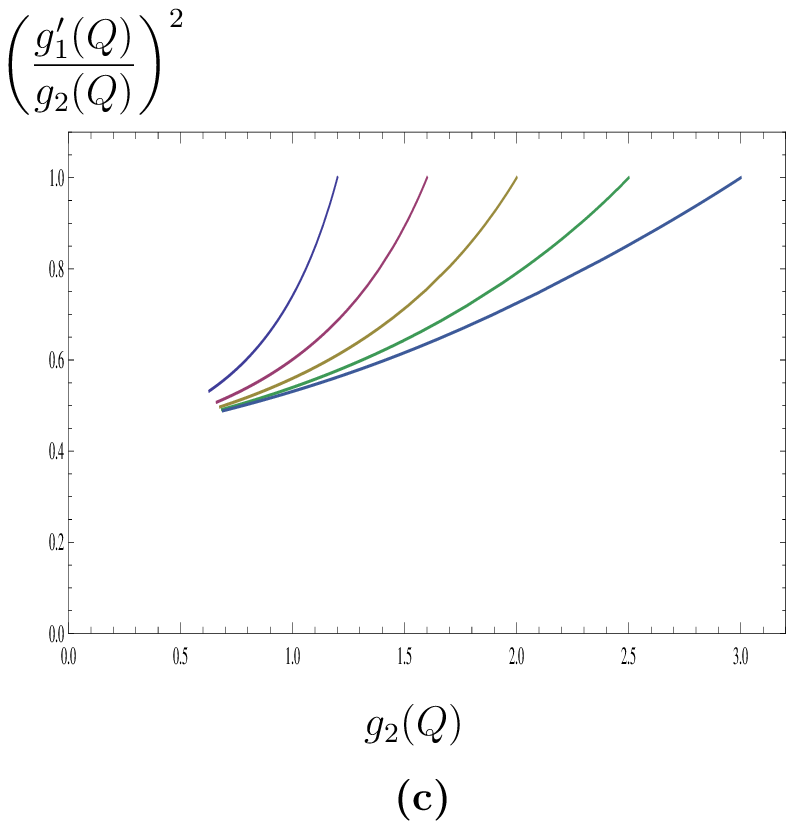}
\includegraphics[width=0.5\textwidth]{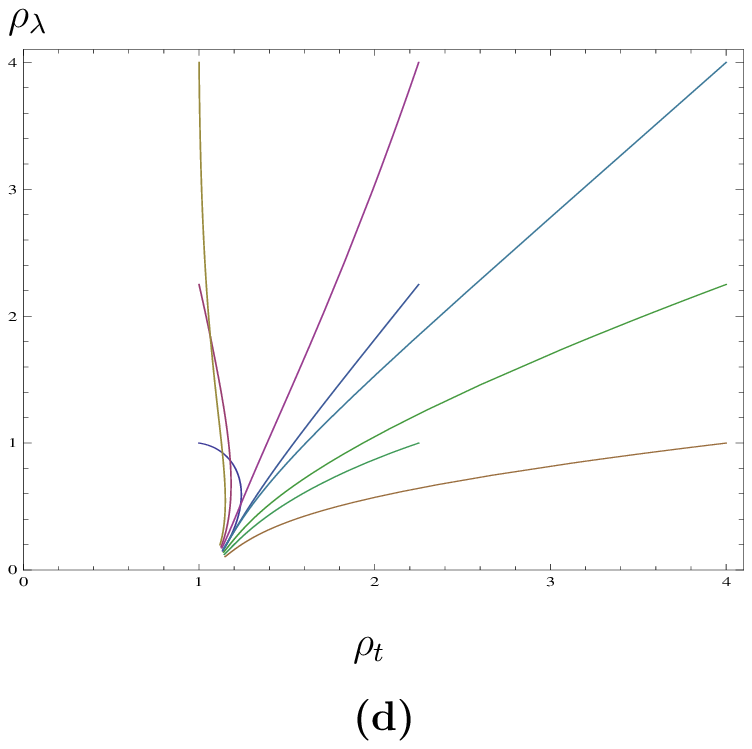}\\
\caption{ Two--loop RG flow of gauge and Yukawa couplings from $Q=M_X$
to EW scale: {\it (a)-(c)} evolution of gauge couplings for
$g_1(M_X)=g'_1(M_X)=g_2(M_X)=g_3(M_X)=h_t(M_X)=\lambda(M_X)=g_0$,
and different values of $g_0$; {\it (d)} running of
Yukawa couplings in the $\rho_{\lambda}-\rho_{t}$ plane for $g_0=1.5$\,.}
\label{fig1}
\end{figure}

In principle, the RG flow of the gauge couplings is affected by the kinetic term mixing. However if the corresponding mixing is small at the
GUT scale $M_X$ then it remains small at any scale below $M_X$. Thus in the leading approximation the kinetic term mixing can be ignored.
Our analysis revealed that the solutions of the two--loop RG equations (RGEs) for the $SU(2)_W$, $U(1)_Y$ and $U(1)_N$ gauge couplings
($g_2$, $g_1$ and $g'_1$) are focused in the infrared region near the quasi-fixed points (see Figs.~1a and 1c) which are rather
close to the measured values of these couplings at the EW scale. On the other hand from Fig.~1b it follows that the convergence
of the solutions for the strong gauge coupling $g_3(Q)$ to the fixed point is quite weak because the corresponding one--loop beta function
vanishes. One can show that the values of $g_i(M_X)=g_0\simeq 1.5$ lead to $g_i(M_Z)$ which are rather close to the measured central
values of these couplings at the EW scale.

As $h_{t}(M_X)$ and $\lambda(M_X)$ grow, the region at the EW scale in which the solutions of the RGEs for $\rho_t$ and $\rho_{\lambda}$
are concentrated shrinks drastically. Thus the corresponding solutions are focused near the quasi--fixed point. This follows from Fig.~1d. 
In this case the quasi--fixed point is an intersection point of the invariant and quasi--fixed lines \cite{Nevzorov:2001vj}--\cite{Nevzorov:2002ub}. 
In the two--loop approximation the coordinates of the quasi--fixed point are $\rho_t=1.16$ and $\rho_{\lambda}=0.14$. 
The quasi--fixed point solution corresponds to $\tan\beta=\frac{v_2}{v_1} \simeq 1$. Our estimations show that for 
$1.5\lesssim h_t(M_X),\,\lambda(M_X)\lesssim 3$ two--loop upper bound on the lightest Higgs boson mass varies
between $120-127\,\mbox{GeV}$ \cite{Nevzorov:2013ixa}.

It is worth noting that for $\tan\beta \simeq 1$ the solutions with $125-126\,\mbox{GeV}$ SM--like Higgs boson can be obtained
only if $\lambda\gtrsim g'_1$ at the EW scale which leads to extremely hierarchical structure of the Higgs spectrum \cite{King:2005jy}.
In this case the qualitative pattern of the Higgs spectrum is rather similar to the one that arises in the Peccei-Quinn (PQ) symmetric 
NMSSM in which the heaviest CP--even, CP--odd and charged states are almost degenerate and much heavier than the lightest 
and second lightest CP-even Higgs bosons \cite{Miller:2003ay}. Since the second lightest CP--even Higgs state and
$Z'$ boson have approximately the same masses and $M_{Z'}\gtrsim 2.5\,\mbox{TeV}$ the heaviest Higgs boson masses lie beyond
the multi TeV range and the mass matrix of the CP--even Higgs sector can be diagonalized using the perturbation theory
\cite{Miller:2003ay}--\cite{Nevzorov:2001um}. In this limit the lightest CP--even Higgs boson is the analogue of the SM Higgs
field. All other Higgs states are so heavy that they can not be discovered at the LHC. Extremely hierarchical structure of the Higgs
spectrum also implies that all phenomenologically viable scenarios associated with the quasi--fixed point are very fine--tuned.

\section{Dark matter and exotic Higgs decays}

The fermionic components of the Higgs--like and SM singlet superfields, which are $\tilde{Z}^{H}_2$ odd, form a set of inert neutralino
and chargino states. Using the method proposed in \cite{Hesselbach:2007te} one can show that there are theoretical upper bounds on
the masses of the lightest and second lightest inert neutralino states  ($\tilde{H}^0_1$ and $\tilde{H}^0_2$) \cite{Hall:2010ix}--\cite{Hall:2010ny}.
Their masses do not exceed $60-65\,\mbox{GeV}$ \cite{Hall:2010ix}. Therefore these states, which are predominantly inert singlinos, tend to be
the lightest and next--to--lightest SUSY particles (LSP and NLSP). In the simplest phenomenologically viable scenarios LSP is considerably lighter
than $1\,\mbox{eV}$ and form hot dark matter in the Universe. The existence of very light neutral fermions in the particle spectrum
may lead to some interesting implications for the neutrino physics (see, for example \cite{Frere:1996gb}). Because LSP is so light it gives only
minor contribution to the dark matter density. At the same time the conservation of the $Z^{M}_2$ and $\tilde{Z}^{H}_2$ symmetries
ensures that the lightest ordinary neutralino is also stable and may account for all or some of the observed cold dark matter density.

The NLSP with the GeV scale mass gives rise to the exotic decays of the SM--like Higgs boson $h_1\to \tilde{H}^0_2\tilde{H}^0_2$.
The couplings of the lightest Higgs state to the LSP and NLSP are determined by their masses. Because $\tilde{H}^0_1$ is extremely light
it does not affect Higgs phenomenology. On the other hand the branching ratio of the nonstandard decays of the SM--like Higgs boson into
a pair of the NLSP states, i.e. $h_1\to \tilde{H}^0_2\tilde{H}^0_2$, can be substantial if NLSP has a mass $m_{\tilde{H}^0_{2}}$ of order
of the $b$--quark mass $m_b$. Nonetheless the couplings of $\tilde{H}^0_1$ and $\tilde{H}^0_2$ to the $Z$--boson and other SM particles
can be negligibly small because of the inert singlino admixture in these states \cite{Nevzorov:2013tta}. As a consequence these states could
escape detection at former and present experiments.

After being produced the NLSP sequentially decay into the LSP and fermion--antifermion pairs via virtual $Z$. Thus the exotic decays of the lightest
CP--even Higgs state discussed above lead to two fermion--antifermion pairs and missing energy in the final state. However since $\tilde{H}^0_2$
tend to be longlived particle it decays outside the detectors resulting in the invisible decays of $h_1$. If $m_{\tilde{H}^0_{2}} \gg m_b(m_{h_1})$
the lightest Higgs boson decays mainly into $\tilde{H}^0_2\tilde{H}^0_2$ resulting in the strong suppression of the branching ratios for the decays
of $h_1$ into SM particles. To avoid such suppression we restrict our consideration to the GeV scale masses of $\tilde{H}^0_2$.  In our analysis we
require that the NLSP decays before BBN, i.e. its lifetime is shorter than $1\,\mbox{sec}$. This requirement rules out too light $\tilde{H}^0_2$
because $\tau_{\tilde{H}^0_{2}}\sim 1/(m_{\tilde{H}^0_{2}}^5)$. One can easily find that it is rather problematic to satisfy this
restriction for $m_{\tilde{H}^0_{2}}\lesssim 100\,\mbox{MeV}$. The numerical analysis indicates that
the branching ratio associated with the decays $h_1\to \tilde{H}^0_2\tilde{H}^0_2$ can be as large as
20-30\% if $\tilde{H}^0_2$ is heavier than 2.5 GeV  \cite{Nevzorov:2013tta}. When $\tilde{H}^0_2$ is lighter than 0.5 GeV this branching
ratio can be as small as $10^{-3}-10^{-4}$ \cite{Nevzorov:2013tta}.

\acknowledgments
This work was supported by the University of Adelaide and the Australian Research Council through the ARC
Center of Excellence in Particle Physics at the Terascale.

\end{document}